\documentclass[runningheads]{llncs}
\usepackage[T1]{fontenc}
\usepackage{graphicx}
\usepackage{booktabs}
\usepackage[misc]{ifsym}
\newcommand{\corr}{(\Letter)}
\usepackage{mwe}

\usepackage[table,svgnames]{xcolor}
\usepackage{caption}
\usepackage{array}
\usepackage{booktabs}
\usepackage{hyperref}
\usepackage{amsmath,amssymb,amsfonts}
\usepackage{algorithmic}
\usepackage{graphicx}
\usepackage{xcolor}
\usepackage{glossaries}
\usepackage{siunitx}

\newacronym{ann}{ANN}{Artificial Neural Network}
\newacronym{cuba}{CuBa}{Current-Based}
\newacronym{cuba-lif}{CuBa-LIF}{Current-Based LIF}
\newacronym{dl}{DL}{Deep Learning}
\newacronym{edp}{EDP}{Energy-Delay Product}
\newacronym{fpga}{FPGA}{Field Programmable Gate Array}
\newacronym{hpo}{HPO}{hyperparameter optimization}
\newacronym{iot}{IoT}{Internet of Things}
\newacronym{kpi}{KPI}{Key Performance Indicator}
\newacronym{lif}{LIF}{Leaky Integrate-and-Fire}
\newacronym{ml}{ML}{Machine Learning}
\newacronym{mpu}{MPU}{microprocessor unit}
\newacronym{rnn}{RNN}{Recurrent Neural Network}
\newacronym{rsnn}{RSNN}{Recurrent SNN}
\newacronym{snn}{SNN}{Spiking Neural Network}

\begin{document}

\title{Heterogeneous SoC integrating an open-source recurrent SNN accelerator for neuromorphic edge computing on FPGA}




\author{Michelangelo Barocci\orcidID{0009-0006-7894-5006}\inst{1} \corr \and
Vittorio Fra\orcidID{0000-0001-9175-2838}\inst{2} \and
Enrico Macii\orcidID{0000-0001-9046-5618}\inst{2} \and 
Gianvito Urgese\orcidID{0000-0003-2672-7593}\inst{2}
}
\authorrunning{M. Barocci et al.}

\institute{
Department of Computer and Control Engineering, \\Politecnico di Torino, Turin, Italy 
\and
Interuniversity Department of Regional and Urban Studies and Planning, \\Politecnico di Torino, Turin, Italy\\ 
\email{\{name.surname\}@polito.it}
}


\titlerunning{Heterogeneous SoC for neuromorphic edge computing on FPGA}

\maketitle              
\begin{abstract}
The growing popularity of Spiking Neural Networks (SNNs) and their applications has led to a significant fast-paced increase of neuromorphic architectures capable of mimicking the spike-based data processing typical of biological neurons. 
The efficient power consumption and parallel computing capabilities of the SNNs lead researchers towards the development of digital accelerators, which exploit such features to bring fast and low-power computation on edge devices. The spread of digital neuromorphic hardware however is slowed down by the prohibitive costs that the silicon tape out of circuits brings, that's why targeting Field Programmable Gate Arrays (FPGAs) could represent a viable alternative, offering a flexible and cost-effective platform for implementing digital neuromorphic systems and helping the spread of open-source hardware designs.
In this work we present an heterogeneous System-on-Chip (SoC) where the operations of ReckOn, a Recurrent SNN accelerator, are managed through the integration with traditional processors. These include the RISC-V-based, open-source microcontroller X-HEEP and the ARM processor featured in Zynq Ultrascale systems.
We validate our design by reproducing the classification results through the implementation on FPGA of the taped-out version of ReckOn in order to check the equivalence of the accuracy and the characteristics in terms of physical implementation. In a second set of experiments, we evaluate the online learning capability of the solution in classifying a subset of the Braille digit dataset recently used to compare neuromorphic frameworks and platforms.

\end{abstract}

\section{Introduction}
Since when it has appeared about forty years ago, the term `neuromorphic' is used to identify systems and models whose design rely on the architecture and working principle of biological nervous systems~\cite{indiveri2011neuromorphic}. The computing and engineering approaches driven by such inspiration therefore aim at emulating the human brain and its functionalities~\cite{markovic2020physics,bartolozzi2022embodied,schmidgall2024brain}.
In the domain of \gls{dl}, the so-called third-generation of \glspl{ann}, namely the \glspl{snn}, gives shape to this brain-inspired paradigm by embedding bio-plausible neuron models to perform computation through the binary counterpart of biological action potentials. 
Referred to as spikes, the latter represent the neural response to a given input, and, in the simplest case, they are generated when an internal state modelling the neuronal membrane potential exceeds a threshold. Such spikes are transmitted among neurons through synapses in the form of weighted sums~\cite{Maass1997networks}.
From the hardware perspective, \glspl{snn} can be deployed with tremendous advantages in terms of energy saving on dedicated chips~\cite{Davies2018loihi,Mayr2019spinnaker,Orchard2021efficient,muller2022braille,Bos2023sub,Pedersen2023neuromorphic} that effectively exploit asynchronous and sparse computation. Nonetheless, the limited accessibility of such platforms, coupled with the still poor availability of other event-based devices for on-edge applications, induces to explore alternative hardware solutions relying, for instance, on tailored design of \gls{snn} accelerators through \glspl{fpga}~\cite{siddique2023low,linaresbarranco2024adaptive,carpegna2024spiker,barocci2023review,pham2021survey}.
Due to the fact that standalone \gls{snn} accelerators perform a highly specific and task-focused set of operations, their configuration often has to undergo the supervision of external controllers \cite{schuman2022opportunities,urgese2023powering}. Solutions based on controllers that operate far from the accelerator are not ideal, since they are not optimized from the point of view of energy efficiency, which is a crucial aspect in edge applications. Having a local controller that manages all the operations and communications is the preferred way for developing neuromorphic applications.
In this paper we present the configuration of an open-source Recurrent SNN accelerator, ReckOn~\cite{frenkel2022reckon}, performed through (1) an open-source microcontroller based on the RISC-V Instruction Set, X-HEEP~\cite{machetti2024xheep}, synthesized on a XC7Z020 FPGA and (2) the ARM processor available in the Zynq SoCs typical of Xilinx FPGA development boards. In the first setup, we used the SPI peripheral of X-HEEP to configure the accelerator, while in the second setup the ARM processor running Linux controls the operations of the accelerator that is implemented in the Programmable Logic (PL) part of the SoC through the use of the memory-mapped I/O AXI IPs. 
We tested the two architectures by reproducing the results on the Cue Accumulation dataset available with the source code of ReckOn, achieving a comparable accuracy above 96\% on the test set. 
Lastly, we show the performances of online learning by evaluating how the system is able to recognize Braille digits on a reduced set of classes from the subset used in~\cite{Pedersen2023neuromorphic}. We measured the accuracy on a 3-class subset that includes the digits \textit{A}, \textit{E}, \textit{U}, achieving 90\% on the test set; then on 4 classes, reaching 78.8\% accuracy in the classification of the digits \textit{A}, \textit{E}, \textit{U}, \textit{Space} and 60\% of digits \textit{A}, \textit{E}, \textit{O} \textit{U}. 
\section{Background}
Digital designs in the early development stages require intensive prototyping before undergoing physical implementation like silicon chip tape out.
Reconfigurable platforms powered by FPGAs represent a cost-effective way to prototype digital solutions and to help the spread and accessibility of open-source designs.
Neuromorphic models, which greatly benefit from digital implementations, have been explored in that sense: Trensch and Morrison~\cite{trensch2022HNC} developed a Hybrid Neuromorphic Compute (HNC) node leveraging a Zynq System-on-Chip; Clair J. et al.~\cite{clair2023spikehard} presented SpikeHard, a neuromorphic accelerator based on an improved architecture of RANC~\cite{mack2021ranc}, where they optimized the neuromorphic cores mapping algorithm to minimize the resources utilization and they integrate the system with the RISC-V core CVA6~\cite{zaruba2019cost}. Neuroflow~\cite{cheung2019neuroflow} and Spiker+~\cite{carpegna2024spiker} bring instead complete training, synthesis and implementation frameworks that assist the developer through all the stages of the SNN definition. Lastly, some projects are also offered via open-source licensing like SENeCA~\cite{2022seneca}, the aforementioned RANC, SyncNN~\cite{2022syncnn} and the target of this work, ReckOn.

\subsection{ReckOn neuromorphic accelerator}
ReckOn is a Recurrent SNN accelerator~\cite{frenkel2022reckon} released under open source license on GitHub\footnote{https://github.com/ChFrenkel/ReckOn} . 
The RSNN model that can be simulated comprises up to 256 input and recurrent Leaky Integrate-and-Fire (LIF) neurons and 16 output Leaky Integrate (LI) neurons.
The peculiarity of ReckOn lies in its capability to perform both classification and regression tasks by leveraging the local in space and time learning technique called e-prop~\cite{bellec2020eprop}, allowing on-edge online learning applications.

\begin{figure}[h]
    \begin{center}
        \includegraphics[width=0.8\linewidth]{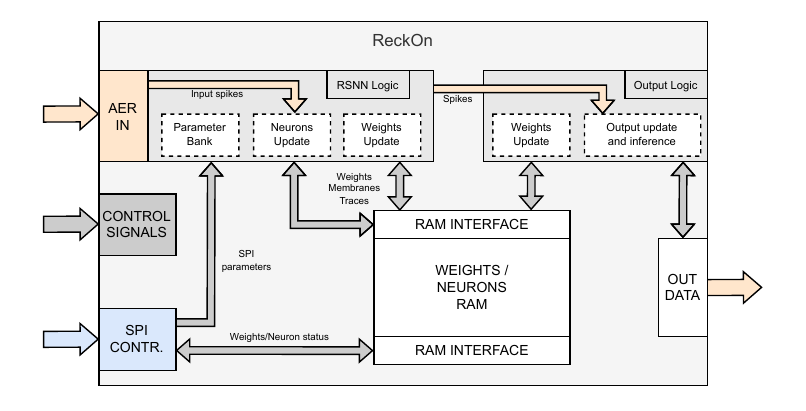}
        \caption{Logic Blocks-level schematic of ReckOn}
        \label{fig:reckon}
        \end{center}
\end{figure}

The internal architecture of ReckOn, reported in Figure~\ref{fig:reckon}, includes a dedicated memory (SRAM) for the storage of the network weights and the hidden neuron status, which includes the membranes' threshold voltages, the leakage factors, the membrane potentials and the eligibility traces that are updated during training. 
The RSNN has direct access to the SRAM blocks to improve the speed of operations: at each timestep, the hidden neurons' membrane voltages and traces are updated sequentially by taking into account the input spikes, leakage factors, and their firing activity.
Input data can be sent to ReckOn encoded as spikes through the Address Event Representation (AER) bus, which consists of an 8-bit address channel and two REQ-ACK  handshake signals. 
ReckOn exposes an SPI interface that allows access to the internal SRAM blocks and the internal parameter bank where the SNN configuration registers are stored to facilitate the integration with traditional systems.
Inference results from ReckOn are provided through an 8-bit bus that reports either the address of the highest output neuron membrane (one time per sample) or a sequence of the output layer's membrane potentials (at each timestep) in case of a classification or regression task, respectively.

\subsection{X-HEEP}
X-HEEP~\cite{machetti2024xheep} is an FPGA-synthesizable open source\footnote{https://github.com/esl-epfl/x-heep} microcontroller based on RISC-V cores such as the \texttt{CV32E20}, \texttt{CV32E40P} and the \texttt{CV32E40X}.
It has been conceived to offer flexible support for external accelerators and to be extended with custom IPs through dedicated interfaces such as the \texttt{CORE-V-XIF} and the \texttt{Extendible Accelerator InterFace (XAIF)} that use the OBI protocol.
Power and resources consumption are some key aspects that are taken into account during the development of X-HEEP, and the system offers customization options that can be applied for tailored applications. It also offers extensive support for the development of software-based applications by supplying HAL and an integration in the PetaLinux ecosystem of Xilinx Ultrascale FPGAs.
\section{Methodology}
We bring ReckOn on the PYNQ-Z2 FPGA by embedding the accelerator as a co-processor in two different systems. The first architecture provides a general purpose solution whose implementation can be deployed on different FPGA platforms, since the microcontroller X-HEEP can be synthesized as a standalone digital system for various hardware targets, the second architecture exploits the flexibility offered by the commercial Xilinx Ultrascale Zynq SoCs, where the Programmable Logic sector is integrated via the AXI bus with the Processing System ARM processor counterpart. The latter solution can be extended only to systems with similar capabilities, since it also includes specific IPs available only through Vivado. 
\subsection{X-HEEP interfaced with ReckOn}
\begin{figure}[h]
    \begin{center}
        \includegraphics[width=0.8\linewidth]{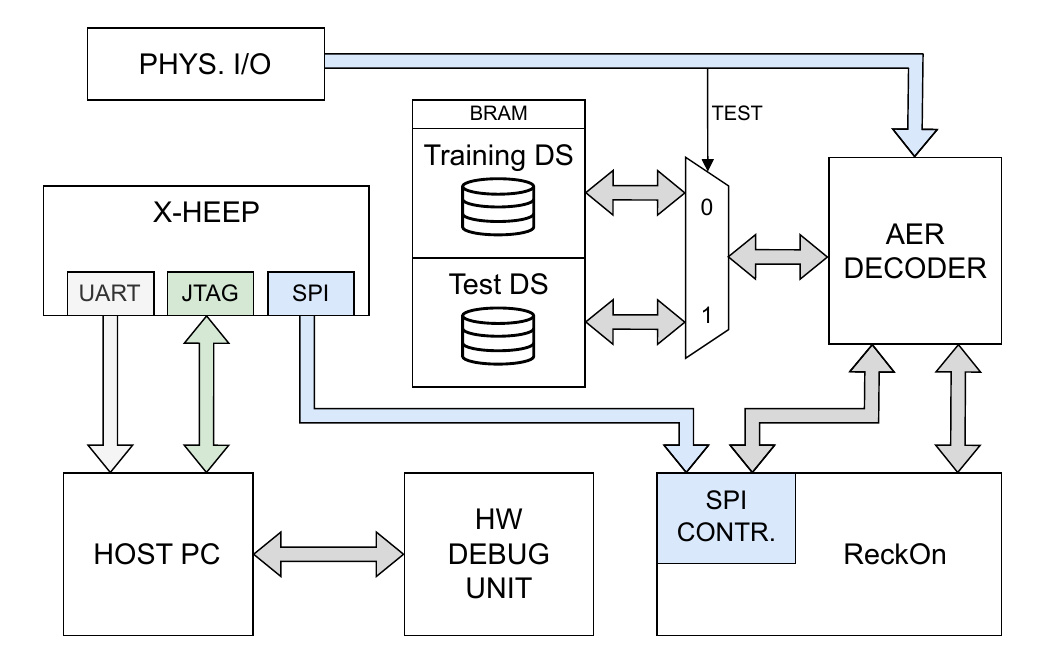}
        \caption{System-level schematic of the architecture based on X-HEEP}
        \label{fig:xheeparch}
        \end{center}
\end{figure}

We implemented ReckOn on FPGA by integrating the accelerator in a system that includes X-HEEP as the main controlling unit. The architecture is shown in Fig.~\ref{fig:xheeparch}. The firmware leverages the Hardware Abstraction Layer (HAL) functions available within X-HEEP that allow to control the peripherals. We used the HAL functions to configure the SPI Host IP to program the parameter bank of the neuromorphic co-processor.

As depicted in Fig.~\ref{fig:xheeparch}, the main subsystems of X-HEEP that are involved include the UART, connected to the host PC for monitoring purposes during the SPI configuration, the JTAG, connected to the PC as well and that is used to deploy and debug the firmware, and the SPI host peripheral, that is used to configure the RSNN parameters of ReckOn. 
We made use of the Internal Logic Analyzer integrated in the Zynq chip to monitor the evolution and the accuracy of the system through the HW debug unit.
Training and validation datasets are loaded during the bitfile writing stage, since they are implemented directly by initializing the BRAMs on the FPGA. 
The interface from the AER decoder towards the RAM is unique, a signal named \texttt{TEST} is used to specify which dataset is being currently read by multiplexing the control signals to the correct memory and the output data bus.
The following paragraph provides an explanation about how data is encoded within memories.

Spikes, labels, and end-of-sample events are stored as 32-bit words, and the information they store varies slightly depending on the type of event that is being targeted. 
The 8 MSBs are dedicated to the type of event: \texttt{0x03} identifies a spike, \texttt{0x02} the label of the sample and \texttt{0x01} the end of the sample. 
Bits from 23 to 12 tell the address of the target neuron for the spike, or the correct label of the current sample.
They are not used at the end of the sample.
Finally, the 12 LSBs indicate the target time tick for the event. This can be interpreted as either the tick at which the spike or the label should be delivered or the final tick of the sample.
\subsection{AER decoder}
\begin{figure}[h]
    \begin{center}
        \includegraphics[width=0.49\linewidth]{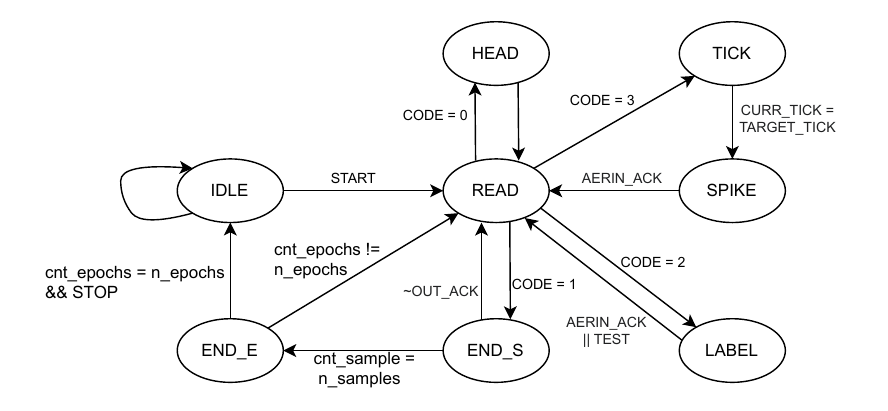}
        \includegraphics[width=0.49\linewidth]{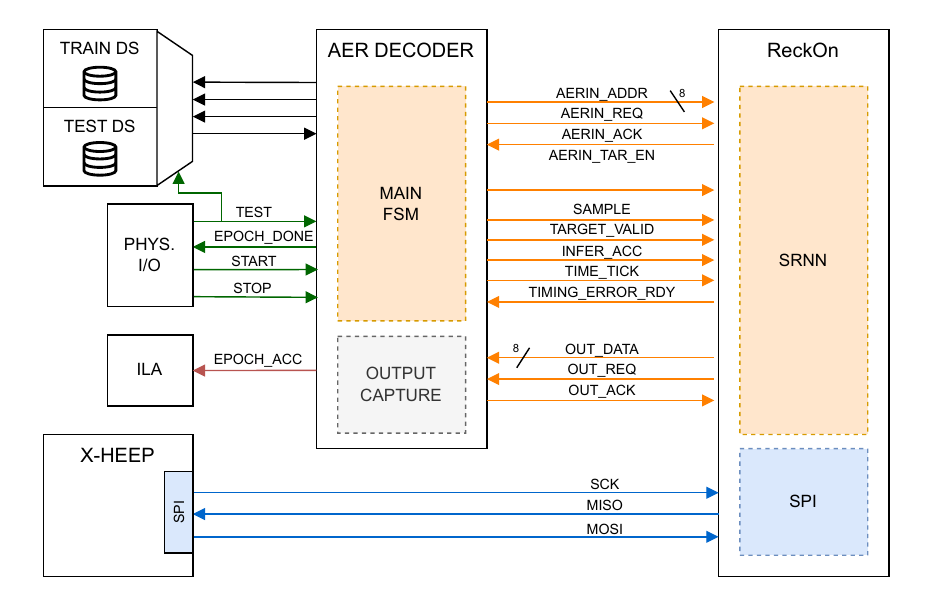}
        \caption{\textbf{Left}: FSM state diagram of the AER decoder in the implementation with X-HEEP. \textbf{Right}: I/O connection of the AER decoder subsystem.}
        \label{fig:fsmX}
        \end{center}
\end{figure}

The binding element between the controllers and the SNN accelerator is the AER decoder, an FSM-based design that is responsible for decoding the 32-bit words stored in the buffer memory into the related input spikes and sending them to ReckOn. 
At the end of each sample, it updates the per-epoch accuracy during training and validation by reading the inferred results generated by the network. 
The integration of the AER decoder within the system can be observed in
Fig.~\ref{fig:xheeparch}.
The sequence of operations performed by the FSM (Fig.~\ref{fig:fsmX} \textbf{Left}) starts from the \texttt{IDLE} state, which waits for the rising edge of the \texttt{START} input signal to assert the \texttt{SAMPLE} signal and start the reading operations from the memory (\texttt{READM} state), then the system increases the current time step counter until the target is reached (\texttt{TICK} state, only in case of a spike), and sends the corresponding spike or label (states \texttt{SPIKE} and \texttt{LABEL}). This set of operations is repeated until the end of each sample (\texttt{END\_S} state), when ReckOn sends the result of the inference after the negation of \texttt{SAMPLE}. 
Once the target number of samples is reached, i.e. the end of an epoch, the accuracy counter is sampled and reset (\texttt{END\_E} state). When the target number of epochs is reached, the system waits for a stop signal to reset itself.

In Fig.~\ref{fig:fsmX} (\textbf{Right}) the schematic of the internal connections involving the AER decoder is depicted. 
The \textit{RAM interface}, controlled by the \texttt{TEST} signal, is used to fetch the training or validation data from the memories, the \textit{control signals} from the Physical I/O subsystem are utilized to initiate/terminate the operations of the accelerator, ascertain whether the AER decoder has completed reading the data or to signal if the current set of samples represents a training or test dataset through \texttt{TEST}, so the AER decoder is able to raise the appropriate signals for learning or inference.
The \texttt{EPOCH\_ACC} counter signal is sampled by the Integrated Logic Analyzer (ILA) which has been synthesized, and it represents the number of correct inferences in the current epoch, the \textit{SPI bus} is used as a mean of communication for the configuration of ReckOn.
Signals related to \textit{ReckOn}, which include the \texttt{AER\_IN} bus, the \texttt{OUT\_DATA} bus, and the control signals to and from the co-processor are on the Right. Input spikes and output inferences employ 8-bit channels and 4-phase handshakes, other signals used to control the operations of ReckOn are the \texttt{SAMPLE} signal that traces the beginning and the end of each sample, the \texttt{TIME\_TICK} signals to communicate each timestep, the \texttt{INFER\_ACC} signal that averages the output membranes to provide the output inference, and the \texttt{TARGET\_VALID} signals which is used to activate the weight updates through the e-prop. Finally, the \texttt{TIMING\_ERROR\_RDY} signal is used by ReckOn to signal the readiness of the system to receive a new timestep.
\subsection{ARM controller interfaced with ReckOn}
\begin{figure}[h]
    \begin{center}
        \includegraphics[width=0.98\linewidth]{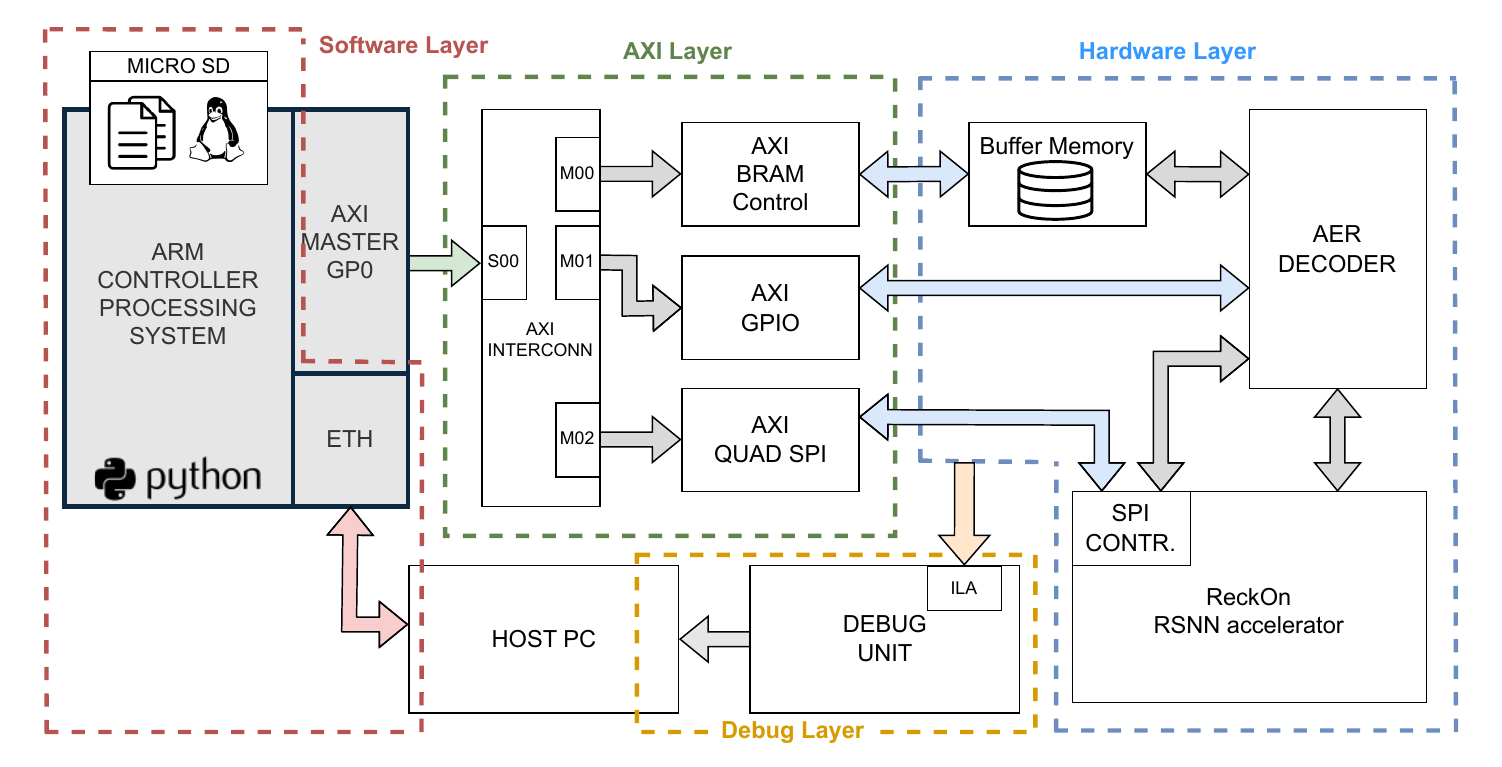}
        \caption{System-level schematic of the ARM-controlled architecture. The dashed lines delimit the perimeter of the four layers.}
        \label{fig:system}
        \end{center}
\end{figure}

\begin{figure}[h]
    \begin{center}
        \includegraphics[width=0.49\linewidth]{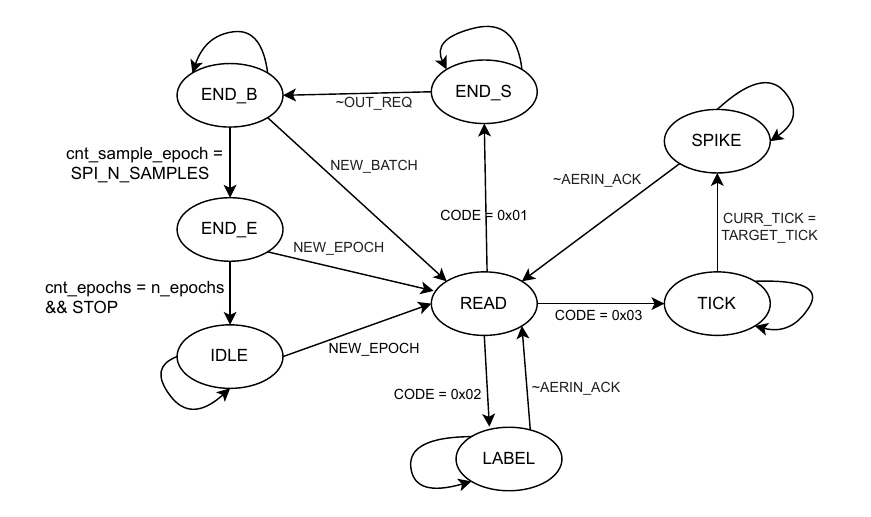}
        \includegraphics[width=0.49\linewidth]{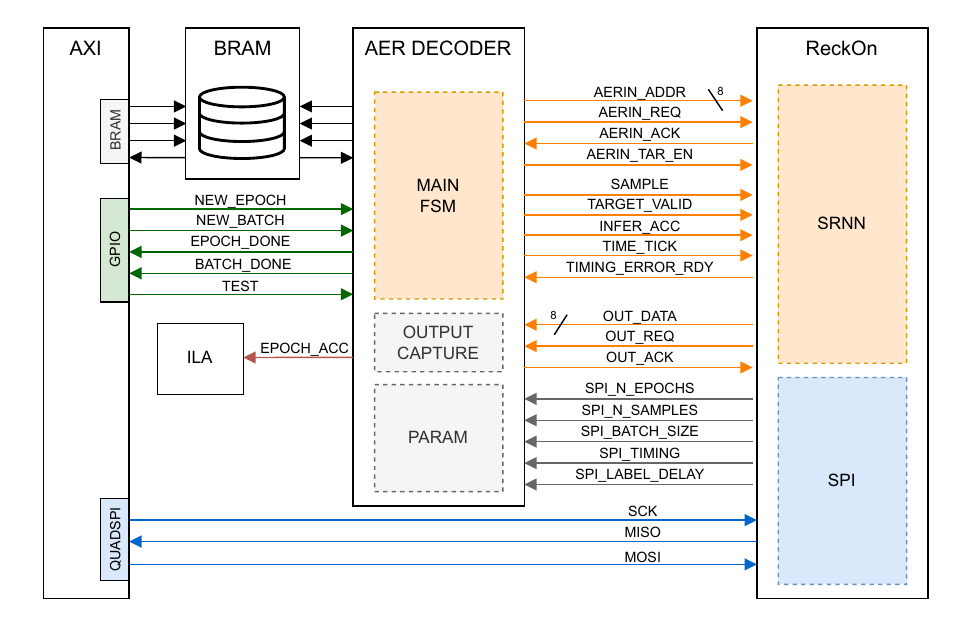}
        \caption{\textbf{Left}: FSM state diagram of the AER decoder in the implementation with the ARM cores, \textbf{Right}: Detailed I/O connection of the AER decoder subsystem with the AXI layer and ReckOn.}
        \label{fig:io}
        \end{center}
\end{figure}

The objective of this second solution is to overcome the constraints imposed by the limited memory of the previous architecture. It is evident that the mechanism of saving all data in the BRAM is not optimal for storing large datasets, such as the Braille dataset that will be presented later. In this second version, we exploited the ARM-based processor available on the Zynq platform to safely store datasets with the operating system and perform periodic batch offload operations on the BRAMs.
The Zynq-7020 chip's processing systems (PS) capabilities are leveraged to program and control the neuromorphic accelerator implemented in the programmable logic (PL) section through the AXI interface. This design can be synthesized only with a specific family of Zynq Ultrascale devices. 
A similar approach has been recently adopted by~\cite{linaresbarranco2024adaptive}.\\
The system is composed of four main layers, as illustrated in Fig.~\ref{fig:system}. These are the \textit{Software layer}, the \textit{AXI layer}, the \textit{Hardware layer}, and the \textit{Debug layer}.
\paragraph{\textit{Software layer}} It revolves around the Jupyter server, which is native to the PetaLinux image of the PYNQ board. 
It is used to access the AXI configuration registers of the different translation IPs. 
Following the initialisation process, during which the accelerator is configured via the SPI interface, the dataset is fed to the accelerator by offloading batches of samples into the buffer BRAM. 
The ARM controller then awaits the assertion of the signals \texttt{BATCH\_DONE} or \texttt{EPOCH\_DONE} by the AER decoder to initiate the offloading of the next batch.
\paragraph{\textit{AXI layer}} It operates on the memory-mapped AXI I/O interface, which is available on the Zynq chip. This interface is the optimal mean of facilitating communication between the PS and PL sides, thanks to the AXI translator bridges available as IPs: the \texttt{AXI-QUADSPI} bridge, the \texttt{AXI-BRAM} bridge and the \texttt{AXI-GPIO} bridge.
Once the IP memory-mapped registers have been written to allow one slave device on the bus and a single-channel MISO/MOSI communication, the \texttt{AXI-QUADSPI} bridge  is employed to perform the initial configuration of ReckOn by reading the necessary data from a text file. 
The samples are offloaded in batches into a shared BRAM memory via the \texttt{AXI-BRAM} bridge. 
Batching is necessary in the case of large datasets that would otherwise require a memory depth exceeding the maximum available BRAM in the XC7Z020 chip. 
Finally, we interact with the system via GPIOs controlled by the \texttt{AXI-GPIO} bridge.
\paragraph{\textit{Hardware layer}} It contains all the HDL architectural components that are added at system-level in the design. It includes all the ReckOn building blocks (SPI slave, SRNN etc.), the FSM-based AER packets decoder and a main wrapper that packages the system with the PS Block design. \\
The previous design of the AER decoder was modified by adding support for the processing of batches of samples. We introduced a new state in the FSM and new control signals, which collectively enhance the flexibility and runtime configuration of the decoder (Fig.~\ref{fig:io} \textbf{Right}).
Signals \texttt{NEW\_EPOCH}, \texttt{NEW\_BATCH}, \texttt{BATCH\_DONE} and \texttt{EPOCH\_DONE} are connected to the GPIO subsystem. As opposed to using physical buttons, this approach simplifies the design by using GPIOs to communicate with the AER packets decoder whenever a batch/epoch has been processed or stored in the memory. Additionally, it allows us to determine whether the current batch is part of a training or validation dataset, as some input signals of ReckOn behave differently in the two cases. Separately, the updated controlling parameters (Fig.~\ref{fig:io} \textbf{Right} in gray) are the \textit{SPI parameters} that are passed to the AER decoder FSM. In this instance, the parameter bank of the SPI slave of ReckOn was expanded with additional parameters that afford greater configuration flexibility during runtime. These include the number of samples per epoch and per batch, the number of epochs, and the delay with which the inference label should be sent (for the delayed supervision task).
In the FSM, the \texttt{HEAD} state was removed, as all configuration is now transmitted via SPI, the state \texttt{END\_B} was introduced, wherein the system enters when the number of processed samples reaches the specified batch size target.
This is indicated by the assertion of the \texttt{BATCH\_DONE} signal, and the system then awaits the controller's action of filling the memory with the subsequent batch and asserting the \texttt{NEW\_BATCH} signal. 
Fig.~\ref{fig:io} (\textbf{Left}) depicts the updated finite state machine (FSM) state diagram of the system.
\paragraph{\textit{Debug layer}} In this layer, the evolution and behavior of the system are monitored with the Internal Logic Analyzer (ILA) feature, which is available for synthesis and can be used directly from the Vivado GUI. Once the set of epochs has been completed and the final testing dataset has been processed, the data is exported and post-processed with a Python script.
\section{Results}
We synthesized our architectures targeting the XC7Z020 programmable chip, demonstrating the feasibility of performing on-the-edge machine learning applications on low-end hardware.
The following subsections report the details of the FPGA resources occupancy. We bring support for both microcontrollers and processors capable of running a standalone Linux-based operating system. On one side, the system allows users to write bare metal code on X-HEEP. On the other side, the system leverages the Python SDK to control the memory-mapped AXI peripheral.
We validate our solutions by comparing the accuracy of the SNN model deployed on our architectures with the ones obtained with the silicon, demonstrating that there is no significant discrepancy.  
Finally, we introduce a new application inspired by a recent study that compared different neuromorphic devices on the Braille digits classification dataset.

\subsection{FPGA resources utilization}
We synthesized the two architectures on a PYNQ-Z2 development board from TUL. 
The board is powered by a Zynq-7020 chip, which offers a powerful environment to test hardware designs thanks to the coexistence of a Dual-Core ARM Cortex-A9 Processors, defined as the Processing System (PS), capable of running Linux and offering extensive SDK for its hardware, and a XC7Z020 FPGA, defined as the Programmable Logic (PL) unit through a multi-purpose AXI bus. The PL can implement digital hardware design that take up to 53200 Look-Up Tables (LUTs), 106400 Flip-Flops, 140 Block RAMs, 220 DSPs and 125 I/O buffers.
\subsubsection{X-HEEP microcontroller - ReckOn implementation}
Ran at \qty{10}{\mega\hertz}, the X-HEEP controlled architecture occupies a total of 45651 LUTs, 145 DSPs, and 94 BRAM tiles. Other resources were allocated for the debug unit, refer to Table \ref{tab:fpgaX} for more details. The presence of initialized BRAMs for the storage of the dataset doubles the ones that are required for the storage of weights and neurons in ReckOn, bringing the total usage at almost 100\%. Also, the coexistence of a microcontroller and its neuromorphic counterpart leaves small room for other possible additional implementations.
\begin{table}[h]
\centering
\rowcolors{2}{lightgray}{white}
\footnotesize
\caption{Zynq-7020 FPGA resource utilization table - with X-HEEP}\label{tab:fpgaX}
\begin{tabular}{lcccc}
                        & \textbf{LUT}$^1$      & \textbf{Flip-Flops}       & \textbf{DSPs}         & \textbf{BRAM}   \\ \hline
ReckOn                  & 31768                 & 3357                      & 144                   & 37              \\ 
AER decoder$^2$         & 131                   & 187                       & 0                     & 41               \\
X-HEEP                  & 13755                 & 17659                     & 1                     & 16              \\
Debug unit$^3$          & 2832                  & 4932                      & 0                     & 5.5            \\
Total (\% Utiliz.) \ \  & \ \ 48483 (91.1\%)\ \ & \ \ 26125 (24.55\%) \ \   & \ \ 145 (65.9 \%)\ \ & \ \ 99.5 (71.07\%) \ \ \\
Total available         & 53200                 & 106400                    & 220                   & 140                  \\ \hline
\multicolumn{5}{l}{
  \begin{minipage}{\textwidth}\vspace{5pt}
    \begin{enumerate}
        \item[$^1$] Includes LUT as logic and LUT as Memory.
        \item[$^2$] Includes also the BRAM resources used to store the two datasets.
        \item[$^3$] The number of LUTs, FFs and BRAM tiles may vary according to the number of debug signals, number of samples and pipe stages.
    \end{enumerate}
  \end{minipage}
}\\
\end{tabular}
\end{table}
\subsubsection*{ARM processor - ReckOn implementation}
Without accounting for the debug unit, the whole architecture took 37528 LUTs, 144 DSPs, and 56 BRAM tiles (more details in Table~\ref{tab:fpga}). We ran the system at a frequency of \qty{15}{\mega\hertz}. With respect to the previous version this design occupies less resources, since the controlling element is already implemented physically in the chip, allowing extra space for additional IPs, like the AXI bridges we make use of. By reducing the amount of BRAMs necessary for the storage of data, it was possible to allocate more resources to the debug unit.
\begin{table}[h]
\centering
\rowcolors{2}{lightgray}{white}
\footnotesize
\caption{Zynq-7020 FPGA resource utilization table - with the ARM cores}\label{tab:fpga}
\begin{tabular}{lcccc}
                        & \textbf{LUT}$^1$      & \textbf{Flip-Flops}       & \textbf{DSPs}         & \textbf{BRAM}   \\ \hline
ReckOn                  & 33764                 & 3342                      & 144                   & 37              \\ 
AER decoder             & 136                   & 186                       & 0                     & 0               \\
PS$^2$                  & 3628                  & 4799                      & 0                     & 19              \\
Debug unit$^3$          & 3804                  & 6633                      & 0                     & 64.5            \\
Total (\% Utiliz.) \ \  & \ \ 41335 (77.7\%)\ \ & \ \ 14977 (14.08\%) \ \   & \ \ 144 (65.45 \%)\ \ & \ \ 120.5 (86.07\%) \ \ \\
Total available         & 53200                 & 106400                    & 220                   & 140                  \\ \hline
\multicolumn{5}{l}{
  \begin{minipage}{\textwidth}\vspace{5pt}
    \begin{enumerate}
        \item[$^1$] Includes LUT as logic and LUT as Memory.
        \item[$^2$] Includes also the shared BRAM resources utilization.
        \item[$^3$] The number of LUTs, FFs and BRAM tiles may vary according to the number of debug signals, number of samples and pipe stages.
    \end{enumerate}
  \end{minipage}
}\\
\end{tabular}
\end{table}
\subsection{Binary decision navigation}
\begin{figure}[h]
    \begin{center}
        \includegraphics[width=0.49\linewidth]{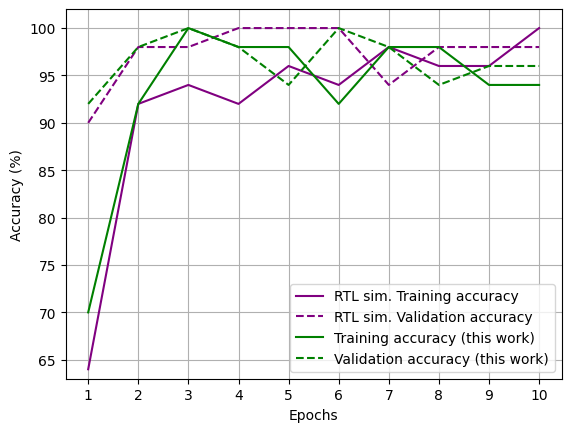}
        \includegraphics[width=0.49\linewidth]{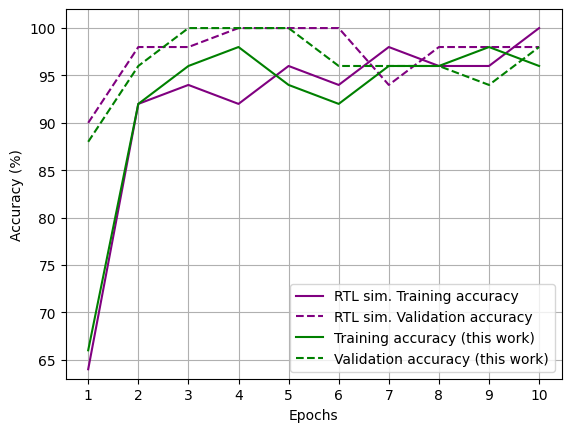}
        \caption{Per-epoch accuracy during training and validation of the binary navigation dataset: X-HEEP implementation to the \textbf{Left} and ARM implementation to the \textbf{Right}. Both results are superimposed to the results of the RL simulation}
        \label{fig:cueacc}
        \end{center}
\end{figure}

We ran the following experiment to demonstrate the equality in terms of accuracy of our FPGA-implemented system and the physical implementation on a silicon chip of ReckOn.
The dataset contains a collection of visual cues that are given to rodents to turn left or right in a controlled environment, and it's used to represent an example of delayed supervision, showing the capabilities of the e-prop learning algorithm in long-term memory SNNs~\cite{bellec2018long}. 
We reproduced the same network, with 40 input neurons, 100 recurrent hidden neurons and 2 output neurons and we performed 10 training and validation epochs on the same 50-samples datasets that is available on the GitHub page of ReckOn\footnote{https://github.com/ChFrenkel/ReckOn} and that is used within the RTL testbench to compare the performances.
Fig.~\ref{fig:cueacc} shows the results obtained using the two configurations, superimposed to the per-epoch accuracy of the RTL simulation. We achieved average 92.4\% (X-HEEP), 92.2\% (ARM) accuracy on the training dataset and 96.8\%, 96.4\% accuracy during validation, while the RTL simulation 92.2\% during training and 97.4\% during validation. The accuracy obtained with the taped out silicon chip, as described in the original paper, is 96.4\%, thus confirming no substantial differences in terms of accuracy.
\newpage
\subsection{Braille digits classification}
We further tested our architecture on the Braille digits dataset~\cite{muller2022braille}, which consists in a set of samples collected by sliding a sensorized fingertip over Braille digits and reading the capacitance values from the 12 separate capacitive sensors over a \qty{60}{\milli\meter} slide. The Braille digits dataset was originally tested in ~\cite{muller2022braille} using a 2-layer Recurrent Spiking Neural Network model on Intel's Loihi
. More recently, the same dataset, reduced to 7 classes, (digits \textit{A} \textit{E} \textit{I} \textit{O} \textit{U} \textit{Y} \textit{Space}) was used to benchmark the RSNN implementations on different neuromorphic commercial hardware exploiting the Neuromorphic Intermediate Representation (NIR)~\cite{Pedersen2023neuromorphic} framework.
With ReckOn, we reduced the dataset to three classes (characters \textit{A} \textit{E} \textit{U}) and we configured the same network used in NIR, i.e. with 12 input neurons and 38 hidden recurrent neurons (reset to zero firing mechanism) and training it on 200 epochs (the number of output neurons depends on the number of classes). 
Later on, we used the same network configuration and perform 200 training epochs using the dataset extended to 4 classes by adding the \textit{Space} and the \textit{O} digits, respectively. 
We divided the full 1400 samples dataset in training (980 samples), validation (280 samples) and test (140 samples) sets, as in NIR. 
We trained the network during 200 epochs and we verified the intermediate results on the validation dataset every 5 epochs. We determined the accuracy at the end of the 200 epochs with the test set.

\subsubsection{3 classes subset}
\begin{figure}[h]
    \begin{center}
        \includegraphics[width=0.49\linewidth]{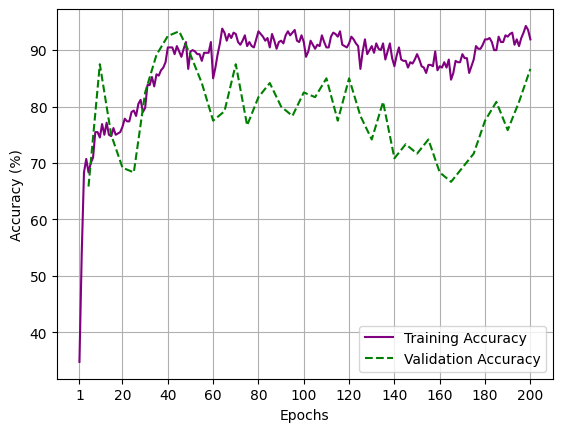}
        \caption{Per-epoch accuracy during training and validation of the RSNN implemented in ReckOn (reset to zero firing mechanism, 38 hidden neurons) with the Braille dataset reduced to three classes, digits \textit{A}, \textit{E}, \textit{U}}
        \label{fig:braille126}
        \end{center}
\end{figure}

Fig.~\ref{fig:braille126} shows the accuracy results during training and validation over 200 epochs on the 3-classes dataset: we achieve an average accuracy of 78.9\% during validation, with the best result (93\%) obtained at epoch 45. The final accuracy is measured at 90\% on the test set using the following hyperparameters: \texttt{0x03F0} as the threshold, \texttt{0x0FE} as the alphas LSBs and \texttt{0x37} as the kappa value.
\subsubsection{4 classes subsets}
\begin{figure}[h]
    \begin{center}
        \includegraphics[width=0.49\linewidth]{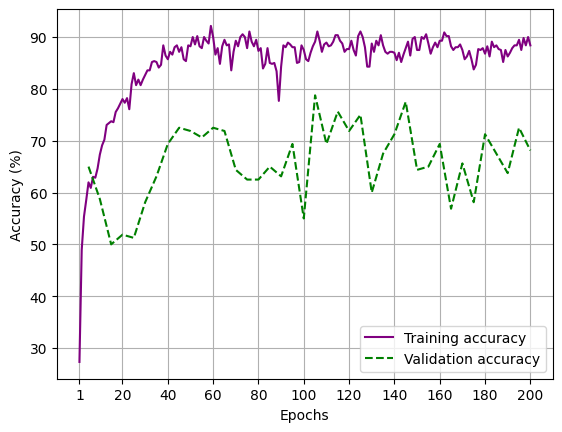}
        \includegraphics[width=0.49\linewidth]{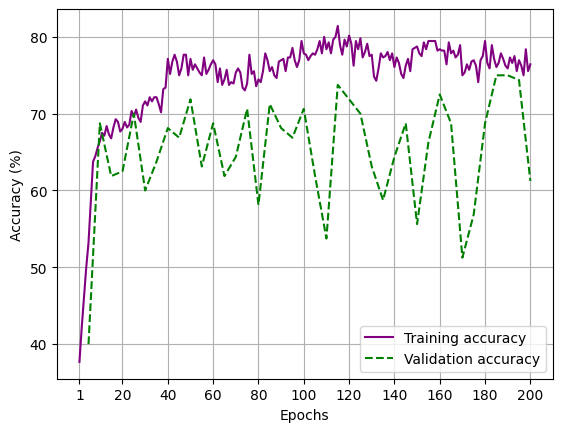}        
        \caption{Per-epoch accuracy during training and validation of the RSNN implemented in ReckOn (reset to zero firing mechanism, 38 hidden neurons) with the Braille dataset reduced to four classes: digits \textit{Space}, \textit{A}, \textit{E}, \textit{U} on the \textbf{Left} and digits \textit{A}, \textit{E}, \textit{O}, \textit{U} on the \textbf{Right}}
        \label{fig:braille4classes_0}
        \end{center}
\end{figure}

The results in Fig.~\ref{fig:braille4classes_0} show the results on the Braille digits dataset reduced to four classes as explained in the section. We achieved a test accuracy of 78.8\% on the dataset containing digits \textit{Space}, \textit{A}, \textit{E}, \textit{U} and 60\% on the dataset containing digits \textit{A}, \textit{E}, \textit{O}, \textit{U} by keeping the same hyperparameters as in the previous configuration.

\section{Conclusion}
We created an interface to control and configure a Recurrent SNN accelerator, ReckOn, creating an heterogeneous ecosystem for the development of neuromorphic embedded applications that leverages either a microcontroller running a firmware, or a dual-core ARM processor running an OS. We integrated ReckOn as a co-processor of X-HEEP and we validate the architecture on FPGA by replicating a use-case previously performed in the silicon chip of ReckOn. Later on, we switched our controlling unit to the ARM-based Processing System of Zynq SoCs running a Linux OS, which commands the accelerator through the AXI bus. This solution allows to execute larger task of edge machine learning by using the internal BRAMs as buffer memories, where the batches of samples can be offloaded temporarily while the whole dataset is safely stored in the internal memory. We validate this architecture by synthesizing the accelerator in the Programmable Logic side of the SoC and by running an online learning task of Braille digits classification.
The latter experiments will be further expanded in future works, which will involve a full comparison with the results obtained with state-of-the-art neuromorphic hardware evaluated in the NIR framework by implementing the same classes in the dataset.

\begin{credits}
\subsubsection{\ackname}
    This research is funded by the European Union - NextGenerationEU Project 3A-ITALY MICS (PE0000004, CUP E13C22001900001, Spoke 6).
    This work is partially supported by the European Commission and the Italian Ministry of Enterprises and Made in Italy (MIMIT) under the KDT ISOLDE project (G.A. 101112274).
    We acknowledge a contribution from the Italian National Recovery and Resilience Plan (NRRP), M4C2, funded by the European Union – NextGenerationEU (Project IR0000011, CUP B51E22000150006, “EBRAINS-Italy”).

\subsubsection{\discintname}
for example: The authors have no competing interests to declare that are
relevant to the content of this article.
\end{credits}

\bibliographystyle{splncs04}
\bibliography{mybibliography}
\end{document}